\documentclass[twocolumn]{svjour3}
\usepackage[utf8]{inputenc}
\usepackage[pdftex]{graphicx}
\graphicspath{{./fig/}}
\usepackage[caption=false]{subfig}
\usepackage{stfloats}
\usepackage{amsmath}
\usepackage{multirow}
\usepackage{fancyvrb}
\usepackage{color}
\usepackage{url}
\usepackage[bookmarks=false]{hyperref}
\usepackage{listings}
\usepackage{todonotes}
\usepackage{booktabs}
\usepackage[english]{babel}

\usepackage{subfig}
\usepackage{booktabs}
\usepackage{gensymb}
\usepackage{pbox}

\date{The final authenticated version is available online at \url{https://doi.org/10.1007/s00450-016-0327-2}.}

\usepackage{color, colortbl}
\definecolor{Gray}{gray}{0.9}
\definecolor{LightGray}{gray}{0.97}

 \usepackage{textcomp}

\newcommand{\figref}[1]{Figure~\ref{#1}}

\newcommand{\tabref}[1]{Table~\ref{#1}}
\newcommand{\secref}[1]{Section~\ref{#1}}

\begin{document}


%
\title{The~Shift~from~Processor~Power~Consumption~to~Perfor\-mance Variations: Fundamental~Implications~at~Scale}

\author{Joseph Schuchart \and Daniel Hackenberg \and Robert Sch\"{o}ne \and Thomas Ilsche \and Ramkumar Nagappan \and Michael K. Patterson}
\institute{
Joseph Schuchart \and Daniel Hackenberg \and Robert Sch\"{o}ne \and Thomas Ilsche
\at
Technische Universit\"{a}t Dresden, Center for Information Services and High Performance Computing (ZIH)\\
01062 Dresden, Germany\\
\email{\{joseph.schuchart, daniel.hackenberg, robert.schoene, thomas.ilsche\}@tu-dresden.de}
\and
Ramkumar Nagappan \and Michael K. Patterson
\at
Intel Corporation \\
2111 NE 25th Avenue, 
Hillsboro, Oregon, USA, 97124 \\
\email{\\\{ramkumar.nagappan, michael.k.patterson\}@intel.com}
}

\maketitle

\begin{abstract}
The Intel Haswell-EP processor generation introduces several major advancements of power control and energy-efficiency features.
For computationally intense applications using advanced vector extension (AVX) instructions, the processor cannot continuously operate at full speed but instead reduces its frequency below the nominal frequency to maintain operations within thermal design power (TDP) limitations.
Moreover, the running average power limitation (RAPL) mechanism to enforce the TDP limitation has changed from a modeling to a measurement approach. 
The combination of these two novelties have significant implications.
Through measurements on an Intel Sandy Bridge-EP cluster, we show that previous generations have sustained homogeneous performance across multiple CPUs and compensated for hardware manufacturing variability through varying power consumption. 
In contrast, our measurements on a Petaflop Haswell system show that this generation exhibits rather homogeneous power consumption limited by the TDP and capped by the improved RAPL while providing inhomogeneous performance under full load.
Since all of these controls are transparent to the user, this behavior is likely to complicate performance analysis tasks and impact tightly coupled parallel applications. 
\end{abstract}

\section{Introduction and Motivation}
\label{sec:intro}
Intel processors code named ``Haswell'' introduce significant architectural changes compared to the previous Intel processor generations.
This includes a distinct frequency used for advanced vector extension (AVX) instructions, P-state transition windows, and a separate uncore frequency~\cite{hackenberg2015energy}.
One implication of the new mechanisms is the shift from focusing on a constant performance across multiple processors to a common power envelope, which is represented by the defined thermal design power (TDP)~\cite{Intel_SpecUpdate_Haswell_EP}.
Previous generations usually exhibited a variable power consumption below the TDP when operating at nominal frequency and therefore exhibited a uniform performance across multiple processors or a whole cluster.
It was only with turbo enabled that the differences in efficiency had an effect on the performance distribution due to the TDP limitations.
The homogeneity in performance has lead to significant differences between processors regarding power dissipation and energy consumption for a given workload while guaranteeing a fairly homogeneous performance pattern.
These efficiency differences between processors of the same model usually stem from hardware manufacturing variability~\cite{variation1,variation2}.

Under the new policy, the processor is limited by its power envelope even at nominal frequency and thus has to scale its computational performance.
This concept is also known as \textit{AVX Turbo Boost}~\cite{Intel_SpecUpdate_Haswell_EP}.
For workloads that utilize AVX instructions, only a defined frequency below the nominal frequency is guaranteed (the \textit{AVX frequency}), while AVX Turbo Boost attempts to achieve higher frequencies if the TDP permits~\cite{intel_avx_whitepaper}.
However, in contrast to traditional turbo, this behavior is transparent to the user even for requested fixed frequencies above the AVX frequency.
This is a dramatic shift as it moves Intel processors from providing homogeneous performance in a cluster towards much more heterogeneous performance characteristics at scale.
However, having predictable and similar performance across a set of CPUs of the same processor model is vital to avoiding artificial imbalances between tightly coupled processes of a parallel application and to ensure reproducibility of performance experiments. 

In this paper, we survey the differences in performance of the new Haswell-EP processor family in a large-scale High Performance Computing (HPC) system described in \secref{sec:test_system}, and present our findings in \secref{sec:perf_scale}. 
In \secref{sec:impact}, we discuss potential effects on applications and performance tuning efforts as well as possible strategies to exploit this heterogeneity. 

\section{Related Work}

We discuss power and performance variations of two generations of Intel processors in this paper.
Wilde et al. describe node-level power variations for more than 260 nodes equipped with AMD Magny Cours executing mprime~\cite{mprime} and Intel Sandy Bridge processors executing FIRESTARTER~\cite{firestarter} in~\cite{wilde}.
They propose to change the batch systems partition to increase the utilization of less power consuming nodes and decrease the utilization of inefficient ones.
Scogland et al. describe power consumption variations over time and between nodes for several systems based on different processors (Intel Nehalem-EP, Nehalem-EX, Sandy Bridge-EP, and AMD FirePro GPUs) with a focus on improving large-scale power measurement methodologies~\cite{Scogland2015}.

However, processor performance variations occur due to power and thermal constraints for certain processors that can be the result of insufficient cooling, artificially low power limits, or hardware over-provisioning.
Starting with Intel Sandy Bridge processors, a power limit below the TDP can be enforced via the running average power limitation (RAPL)~\cite{rapl}.
To estimate the current power consumption, RAPL uses an energy model on Sandy Bridge processors~\cite{rapl_quant} and measurements on Haswell processors~\cite{hackenberg2015energy} to throttle the performance of the processor to achieve the given power limit.
The quality of the RAPL power enforcement in terms of stability, accuracy, settling time and maximum overshoot has been analyzed by Zhang and Hoffman for Sandy Bridge processors~\cite{rapl_quality_constraint}.
Rountree et al. show how an enforced power limit turns power variations into performance variations for Intel Sandy Bridge processors~\cite{rapl_powerbound}.
Pedretti et al. investigate in~\cite{Power_Cap_Cray:2015} the influence of such power limits on power consumption, performance, and energy efficiency of four parallel benchmarks and compare it to the usage of processor P-States.
They describe that an enforced common power cap regulation can limit scalability due to performance variability.

The major changes of the Intel Haswell processor generation with respect to on-chip energy management have been summarized previously~\cite{hackenberg2015energy}.
It was predicted that these changes significantly influence power consumption and performance variations, which is investigated in detail in the following sections.

\section{Test Systems and Measurement Methodology}
\label{sec:test_system}
To demonstrate the power and performance variations at scale, we launch an instance of the LINPACK benchmark on all sockets or nodes, as described below.
All measurements were conducted on the Bullx DLC B710/ B720~\cite{bull_dlc} based system Taurus installed at TU Dresden, for which details are provided in \tabref{tab:system}.
Both the Intel Xeon E5-2690 (Sandy Bridge-EP) and the Intel Xeon E5-2680~v3 (Haswell-EP) partitions contain directly liquid cooled (DLC) dual-socket nodes.
Due to time constraints, not all nodes have been taken into account but the selection of nodes has been unbiased and we believe that the number of nodes presented in \secref{sec:perf_scale} is sufficient for our analyses.  
%

\renewcommand{\arraystretch}{1.2}

\begin{table}[b!]
\caption{Test system description}
\centering
\begin{tabular}{lcc}
\toprule
Processor arch. ~ & ~Sandy Bridge-EP~~ & ~~Haswell-EP~~ \\
\midrule
Intel Xeon &  E5-2690 & E5-2680 v3\\ 
Sockets per node & 2 & 2 \\
Cores per CPU & 8 & 12 \\
Nominal freq. & 2.9\,GHz & 2.5\,GHz \\
Turbo freq. & up to 3.8\,GHz & up to 3.3\,GHz \\
AVX base freq. & n/a & 2.1\,GHz \\
TDP & 135\,W & 120\,W \\
Linux Kernel & \multicolumn{2}{c}{2.6.32-431.23.3.el6} \\
DRAM & \pbox{.2\columnwidth}{\centering 32\,GB DDR3-1600} & \pbox{.2\columnwidth}{\centering 64\,GB DDR4-2133} \\
Number of nodes & 228 & 1328 \\
\bottomrule
\end{tabular}
\label{tab:system}
\end{table}

\begin{figure*}[b]
\center
\hspace{3em}
\subfloat[][Sandy Bridge LINPACK performance]{\includegraphics[width=.85\columnwidth]{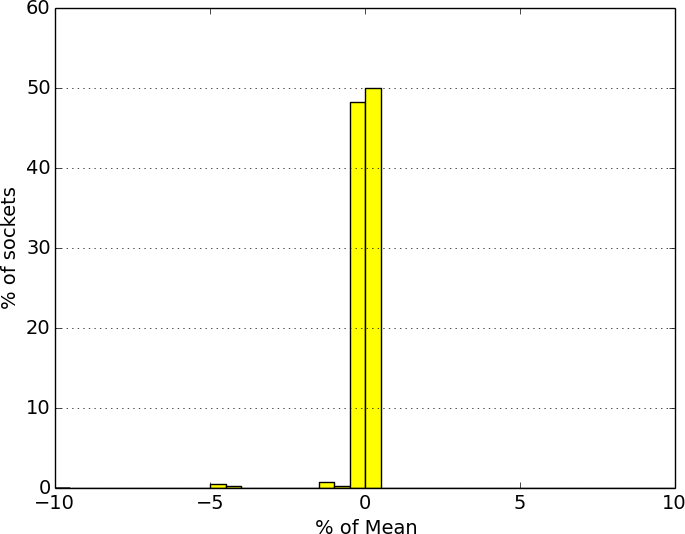} \label{fig:sandy_perf_2900}}
\hfill
\subfloat[][Haswell LINPACK performance]{\includegraphics[width=.85\columnwidth]{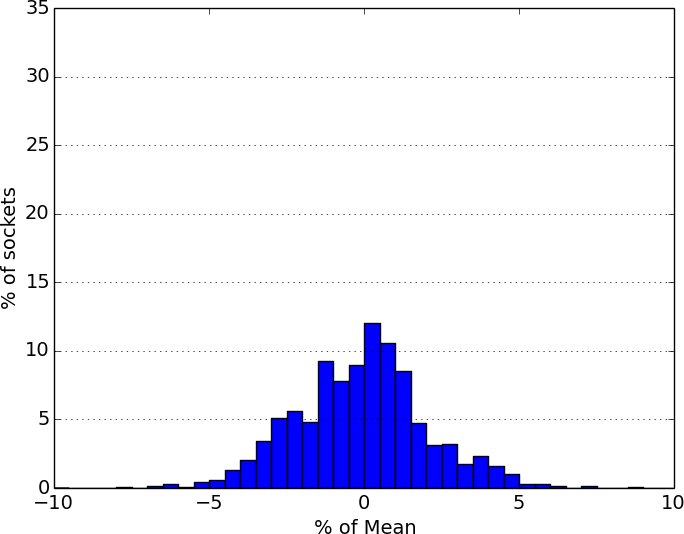} \label{fig:hsw_perf_2500}}
\hspace{3em}

\hspace{3em}
\subfloat[][Sandy Bridge avg. power consumption]{\includegraphics[width=.85\columnwidth]{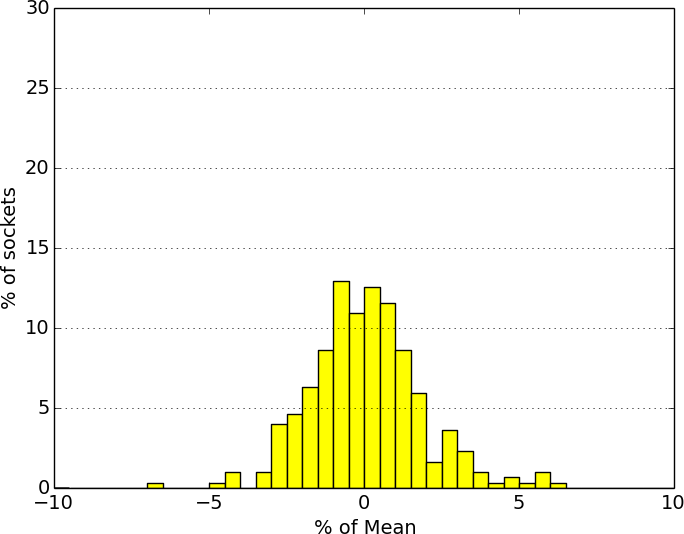} \label{fig:sandy_rapl_2900}}
\hfill
\subfloat[][Haswell avg. power consumption]{\includegraphics[width=.85\columnwidth]{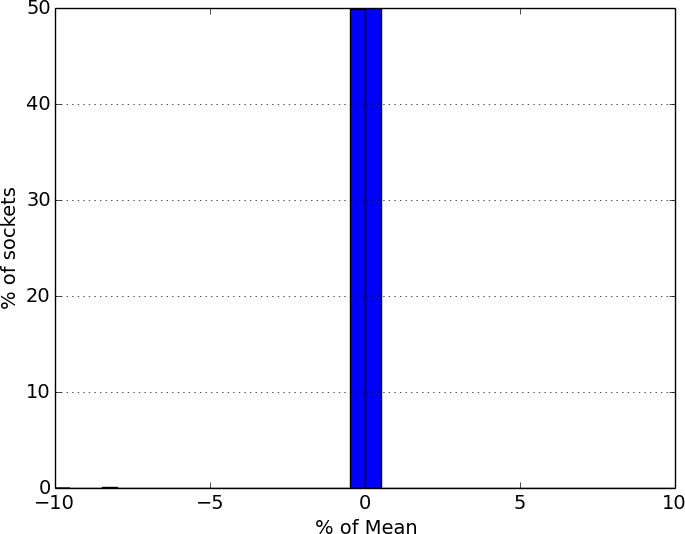} \label{fig:hsw_rapl_2500}}
\hspace{3em}
\caption{Distribution of LINPACK performance and average RAPL package power consumption during the core phase using 412 Sandy Bridge and 1144 Haswell sockets at nominal frequency ($f_{SB} = 2.9$\,GHz, $f_{HSW} = 2.5$\,GHz).
}
\label{fig:sandy_vs_haswell}
\end{figure*}

We use the LINPACK shipped with the Intel MKL and perform ten repetitions of the benchmark.
We take the median performance of these ten runs to be able to rule out possible outliers due to temporary effects.
Similar to the LINPACK performance, we use the measurements of the average power consumption of the run with the median performance. 
We consider the three metrics LINPACK performance, package power as reported by RAPL, and per-node power consumption. 

The LINPACK performance is measured for each processor individually with threads pinned to the respective sockets. 
Two instances per node are launched in parallel to ensure the nodes are under full load, e.g., to rule out potential effects from the cooling system and to prevent changes in package C-states, which may influence the performance of other packages by delaying cache coherence traffic responses.
%
The configuration of each of the LINPACK instances is chosen to utilize nearly half of the available 64\,GB of main memory on the node and to ensure sufficient LINPACK runtime.
Time\-stamped RAPL measurements are taken once a second to detect and handle overflows. 

In addition to the RAPL power, we measure the node power consumption using the HDEEM infrastructure~\cite{hdeem}.
All Haswell nodes are calibrated to facilitate node-level power measurements with 2\,\% accuracy.
The node power measurements using HDEEM are performed with one LINPACK instance per node to account for the spatial granularity of the node-level measurements. 

While the LINPACK benchmark has been critized for not accurately representing real-world HPC workloads, we argue that it is the best choice to easily rank socket performance for compute-bound problems.
The effect of the performance variations presented in this paper are most notable for this class of problems.
We only consider the core phase of the LINPACK for all power measurements on the Haswell nodes.
Due to software restrictions, the exact timestamps of the LINPACK core phase are not available on the Sandy Bridge nodes. 
We therefore use the inner 50\,\% of the whole LINPACK run (starting at 25\% of the runtime) to ensure that only samples from the core phase are used.
The Linux userspace governor is used to repeat the experiments at different core frequencies.


\section{Performance and Power Characteristics}
\label{sec:perf_scale}

\begin{figure*}[b]
\center
\hspace{3em}
\subfloat[][Turbo]{\includegraphics[width=.80\columnwidth]{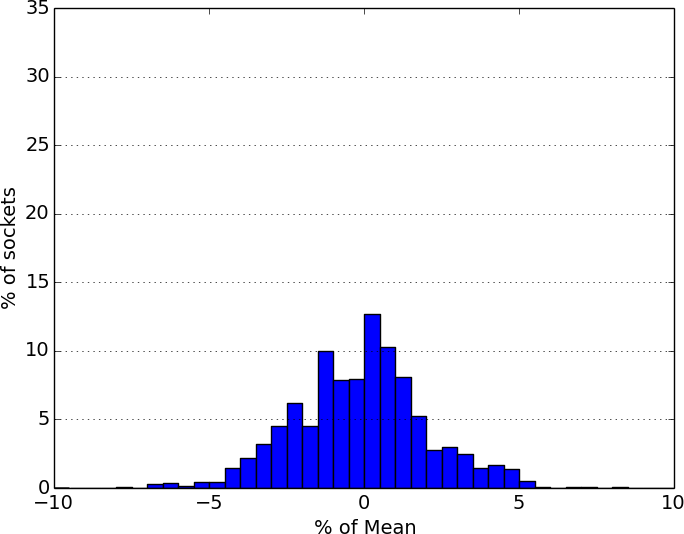}}
\hfill
\subfloat[][2.4\,GHz]{\includegraphics[width=.80\columnwidth]{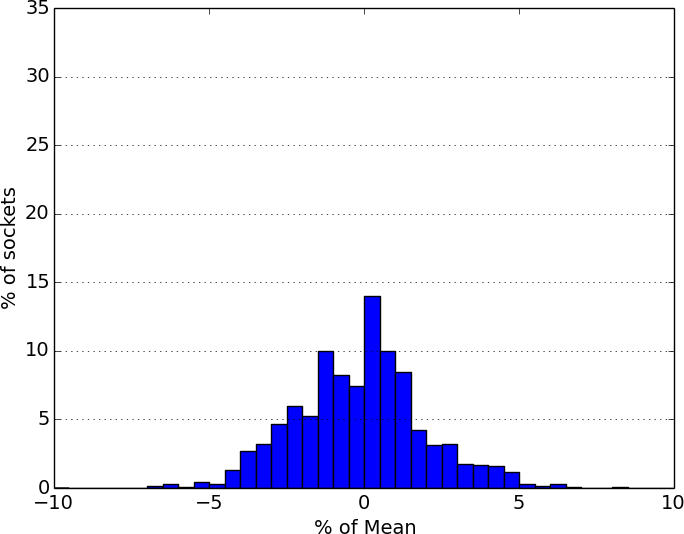}}
\hspace{3em}

\hspace{3em}
\subfloat[][2.2\,GHz]{\includegraphics[width=.80\columnwidth]{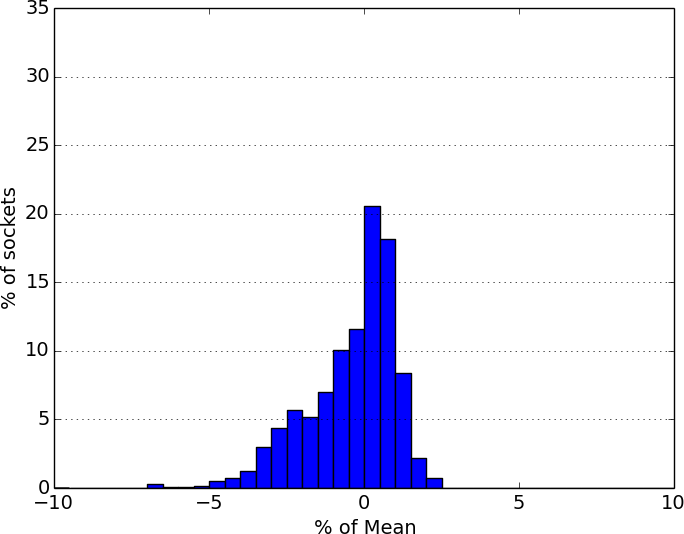}}
\hfill
\subfloat[][2.1\,GHz]{\includegraphics[width=.80\columnwidth]{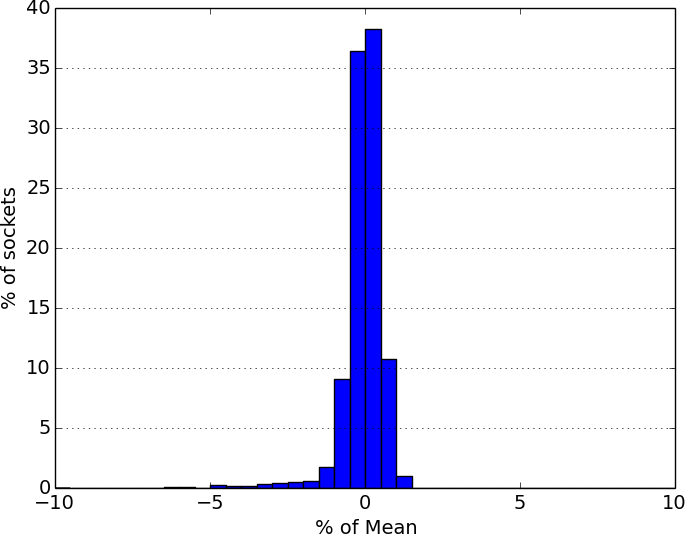}\label{fig:hsw_perf_2100}}
\hspace{3em}
\caption{Per-socket LINPACK performance distribution of 1144 Haswell sockets at different frequencies. Note the diverting y-axis scale in plot (d). At 2.4,GHz and 2.3\,GHz, a similar distribution has been observed as for 2.5\,GHz.}
\label{fig:performance_haswell_rest}
\end{figure*}


The results of our measurements presented in \figref{fig:sandy_perf_2900} indicate that the Sandy Bridge processors show a homogeneous distribution of the performance across all processors under test with the mean performance at 148\,GFlop/s and standard deviation $\sigma = 0.6$\,GFlops/s.
In contrast to that, the performance of the Haswell processors ranges from 355 to 400\,GFlop/s, a variation of more than 10\,\% between the slowest and fastest processor (see \figref{fig:hsw_perf_2500}).
The mean performance is 380\,GFlop/s with $\sigma = 8$\,GFlop/s, constituting a much larger spread compared to Sandy Bridge.

With regards to the average power consumption of the core phase measured through RAPL, the picture is inverted. 
On the one hand, the average power consumption of the Sandy Bridge processors spreads by about $\pm5\,\%$ around the mean of 129\,W ($\sigma = 2.35$\,W).
On the other hand, the power reported by the Haswell processor is highly uniform across all CPUs tested with the average at 119.6\,W ($\sigma = 0.398$\,W).

The uniformity of RAPL measurements on Haswell are as expected considering that the same measurements are used for power capping.
It should be noted that RAPL has only been studied for systematic errors but the absolute accuracy has not been evaluated so far~\cite{rapl_quant,hackenberg2015energy}. 
Inaccuracies in the calibration of the RAPL measurements could contribute to the variations in computational performance across the set of processors.

\subsection{Detailed Performance Analysis}

\begin{figure*}
\center
\hspace{3em}
\subfloat[][Sandy Bridge]{\includegraphics[width=.85\columnwidth]{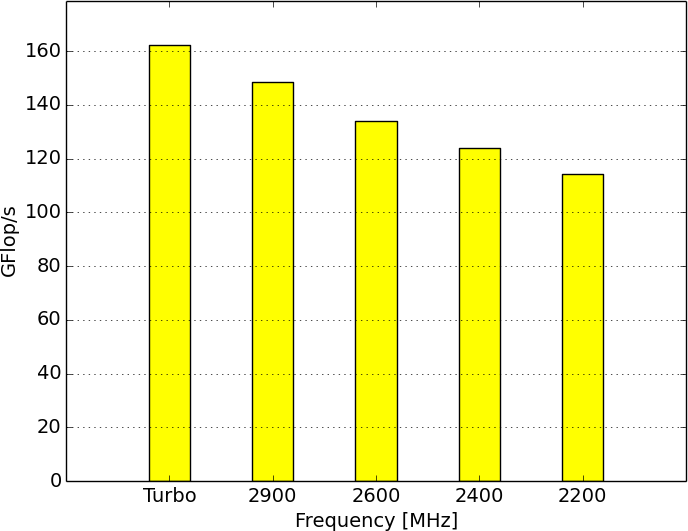}}
\hfill
\subfloat[][Haswell]{\includegraphics[width=.85\columnwidth]{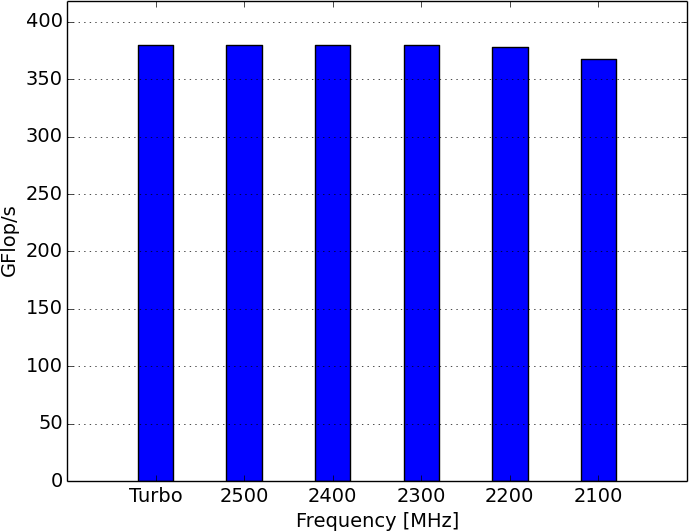}
\hspace{3em}
\label{fig:performance_mean_hsw}}
\caption{Absolute overall mean LINPACK performance per socket for different frequencies.
}
\label{fig:performance_mean}
\end{figure*}

\label{sec:perf_analysis}
For the Sandy Bridge processors under test, the distribution of LINPACK performance was found to be narrow across all measured frequencies, ranging from 2.9\,GHz down to 2.2\,GHz (see \figref{fig:sandy_perf_2900}). 
Only with turbo mode enabled, there is a notable spread in performance around the mean of 162.39 ($\sigma = 1.64$\,GFlop/s, not depicted). 

In contrast to that, the performance distribution pattern of the Haswell processors changes quite significantly at reduced clock speeds as depicted in \figref{fig:performance_haswell_rest}. 
While the distribution remains the same for turbo mode down to 2.3\,GHz, the upper half is cut off starting at 2.2\,GHz.
At 2.1\,GHz and below, the performance distribution is narrow.

\figref{fig:performance_mean} depicts the mean performance over all nodes for different frequencies measured on both processor generations. 
As expected, the performance decreases for lower frequencies on Sandy Bridge processors. 
However, on Haswell processors the average LINPACK performance remains stable from turbo mode down to 2.2\,GHz, with only a minor drop at 2.1\,GHz (see~\figref{fig:performance_mean_hsw}).
This behavior can be explained with the new AVX frequency, which is the minimum guaranteed frequency that can be achieved in the presence of AVX instructions under heavy load. 
This concept has been introduced to ensure that the processors stay within the specified TDP. 
The AVX frequency of our Haswell-EP model is 2.1\,GHz (see~\tabref{tab:system}), which is in line with our experimental results.

\begin{figure}[b]
\center
\includegraphics[width=\columnwidth]{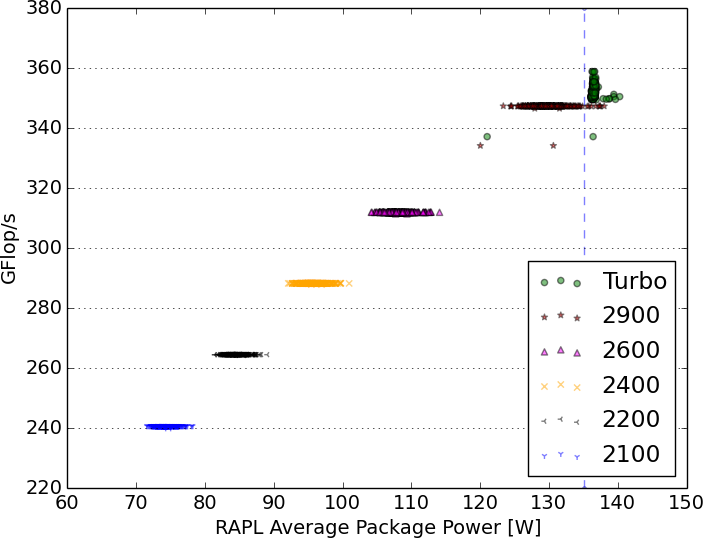}
\caption{Correlation between mean RAPL package power in the core phase and LINPACK performance for different frequencies of the Sandy Bridge processors. The TDP of 135\,W is marked as dashed blue line.}
\label{fig:power_performance_sandy}
\end{figure}

\figref{fig:power_performance_sandy} shows the correlation between the average RAPL package power for the core phase and the achieved LINPACK performance on Sandy Bridge processors. 
The graph reflects the findings described above, the horizontal lines represent the stable performance and varying power dissipation across the set of processors under test. 
The distribution is highly regular along the power axis for all non-turbo frequencies.
If turbo is enabled, a rectangular shape develops with all processors exceeding the specified TDP according to RAPL. 
Most of the processors only slightly exceed the TDP limit and vary by about ten~GFlop/s with similar power consumption (vertical line).
It appears, that the power manager forces a minimum performance (frequency slightly above nominal, horizontal line) for turbo mode regardless of TDP limitations, while enabling a higher performance if the TDP allows it.
It should be noted that some processor even exceed the TDP at nominal frequency.

\begin{figure*}[t]
\center
\subfloat[][Turbo]{\includegraphics[width=.31\textwidth]{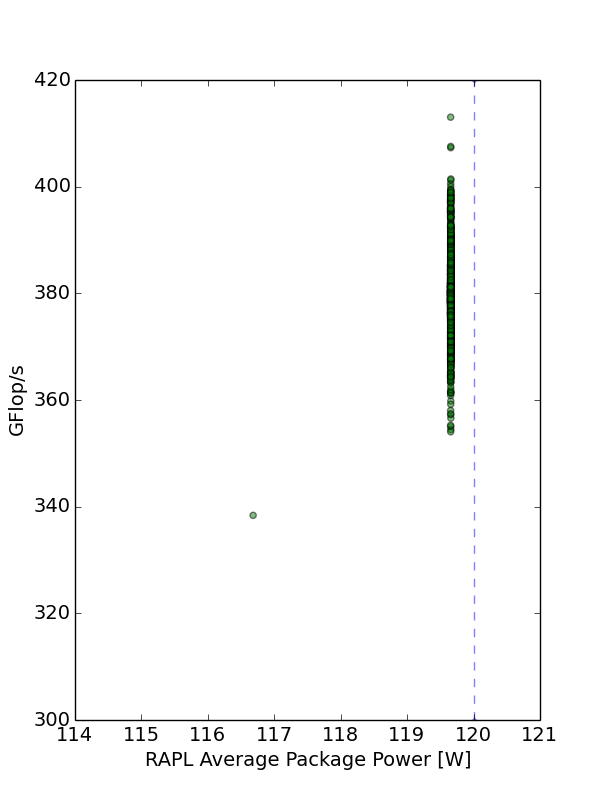}\label{fig:power_flops_haswell_turbo}}
\hfill
%
\hfill
\subfloat[][2.2\,GHz]{\includegraphics[width=.31\textwidth]{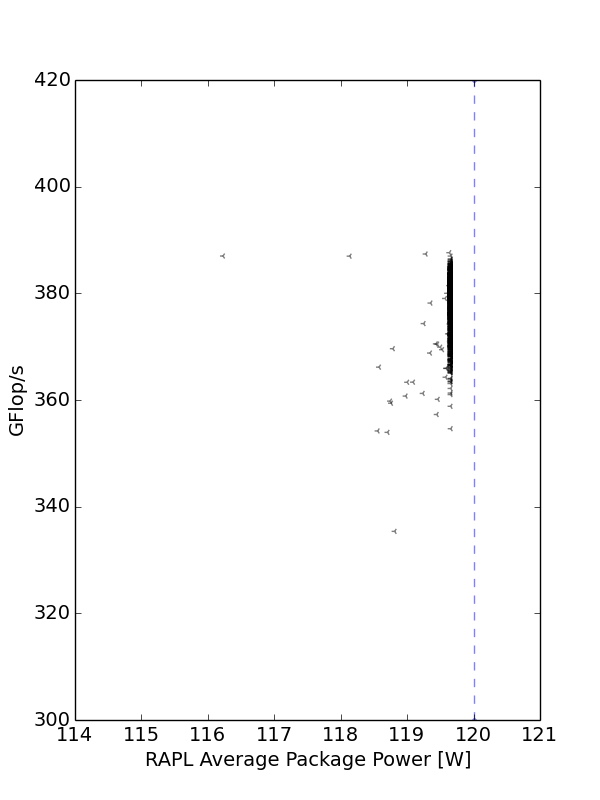}}
\hfill
\subfloat[][2.1\,GHz]{\includegraphics[width=.31\textwidth]{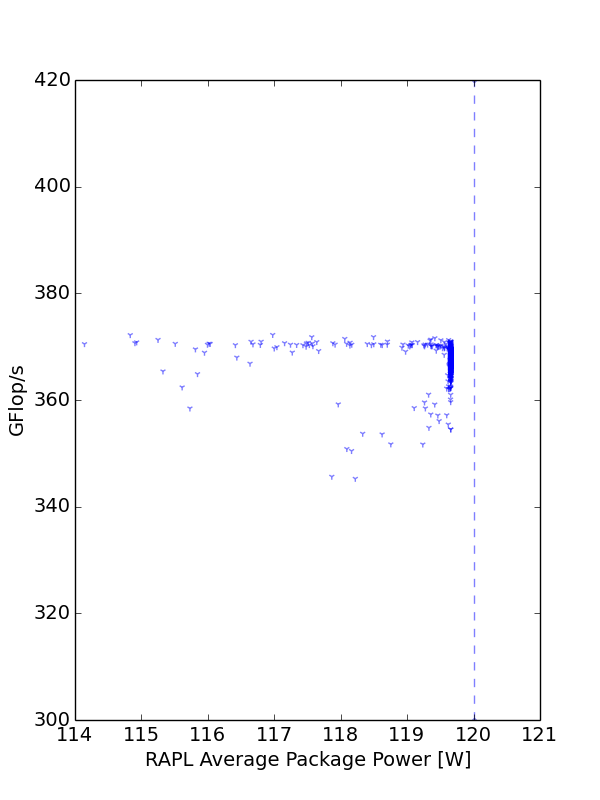}}
\caption{Correlation between mean RAPL package power in the core phase and LINPACK performance for different frequencies of the Haswell processors. The TDP of 120\,W is marked as dashed blue line. Some nodes showed thermal problems, which caused the significant outliers.
}
\label{fig:power_flops_haswell}
\end{figure*}

For Haswell, the correlation between average package power and LINPACK performance is presented in \figref{fig:power_flops_haswell}. 
Here, the clusters are vertical, reflecting a constant power dissipation with varying performance across the set of processors under test. 
At 2.1\,GHz, again a rectangular shape develops, with some processors requiring less power to sustain constant performance. 
These are the processors that exhibit a high performance at higher frequencies and thus are able to reduce their power consumption to match the 2.1\,GHz. 
However, a significant amount of processors are still running at full TDP with their performance below the faster processors. 
In contrast to Sandy Bridge, the TDP is never exceeded in such a long-term scenario.

Some performance outliers can be seen in the plots, e.g., at 340\,GFlop/s in \figref{fig:power_flops_haswell_turbo}, which are likely to be caused by cooling problems as the respective nodes reported high temperature readouts after the runs with some even close to the critical limit. 
While these variations are significant, they should not be considered part of the performance variations discussed in this paper.


\subsection{Node Power Measurements}

\begin{figure*}[t]
\center
\hspace{3em}
\subfloat[][Sandy Bridge]{\includegraphics[width=.85\columnwidth]{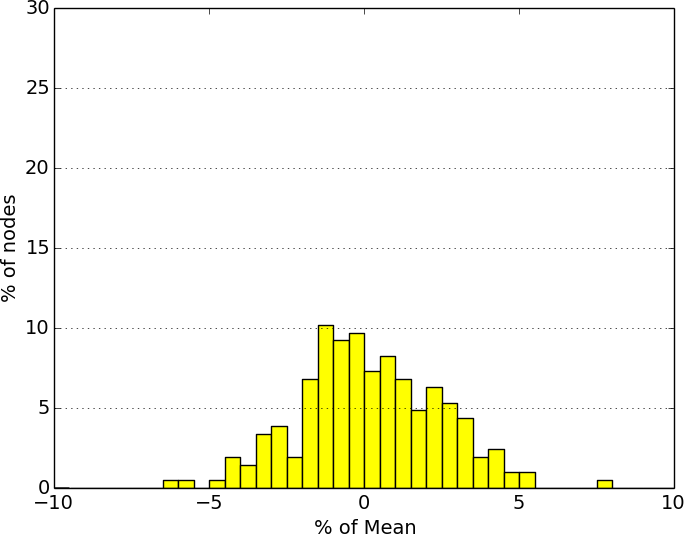}}
\hfill
\subfloat[][Haswell]{\includegraphics[width=.85\columnwidth]{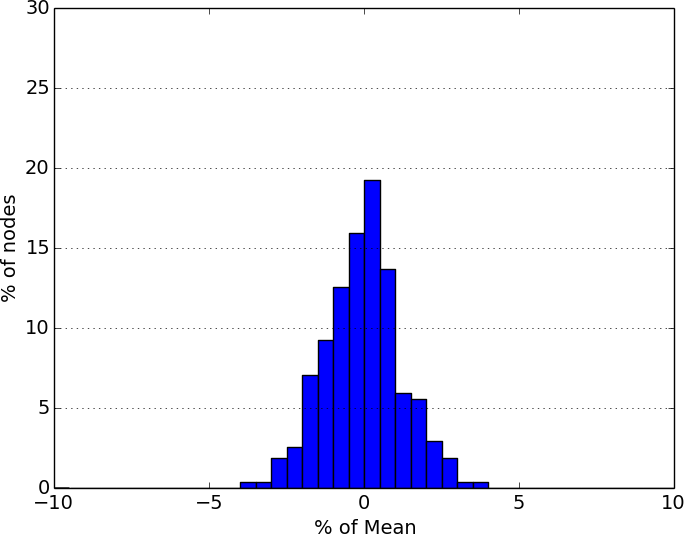}}
\hspace{3em}
\caption{Per-node average power distribution of 151 Sandy Bridge and 270 Haswell dual-socket nodes at nominal frequency for the core phase of the LINPACK run measured with the HDEEM infrastructure.
}
\label{fig:hdeem_spread}
\end{figure*}

\figref{fig:hdeem_spread} shows the distribution of the average power consumption of the used Sandy Bridge and Haswell nodes while running a full-node LINPACK benchmark at nominal frequency. 
In contrast to the performance characteristics on Sandy Bridge processors, the spread of the power consumption across all nodes is far more significant.
Thus, the performance is rather homogeneous at the expense of substantial differences in power consumption due to hardware manufacturing variability, as described in \secref{sec:intro}.

On the Sandy Bridge system, the node power consumption shows a wider spread with only 68\,\% in the range of $\pm2\,\%$ around the mean of 201\,W ($\sigma = 5.1$\,W).
This aligns well with the observations presented in \figref{fig:sandy_rapl_2900}. 
For the Haswell nodes, the distribution shows a smaller spread with about 92\,\% of the nodes in the range of $\pm2\,\%$ around the mean of 342.9\,W ($\sigma = 4.2$\,W).
However, the spread is still more significant then the RAPL measurements presented in \figref{fig:hsw_rapl_2500}, potentially due to the 2\,\% accuracy of HDEEM.  

Given that for both processor generations the node measurements include two sockets as well as memory, network, and local storage, a quantitative comparison with the per-socket RAPL measurements presented in \figref{fig:sandy_vs_haswell} cannot be drawn due to the lack of reliable fine-grained external measurements. 
Nevertheless, the node-power measurements reflect the shift to a uniform power distribution of the Haswell processors. 

\subsection{CPU Temperature Measurements}

\begin{figure}[b]
\center
\includegraphics[width=\columnwidth]{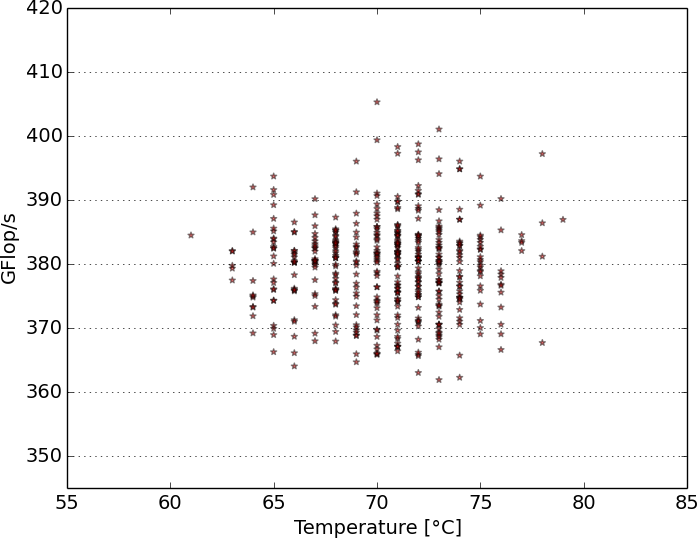}
\caption{Correlation between socket temperature and the LINPACK performance of the Haswell processors at 2.5\,GHz.} 
\label{fig:temp_flops}
\end{figure}

To investigate the impact of the temperature of the Haswell processors on the LINPACK performance, we correlate the temperature of both CPUs in a node after ten runs with the median performance as depicted in \figref{fig:temp_flops}. 
The scatter plot shows no strong correlation between the two variables.
We assume that the significant difference in temperature between the coolest and hottest CPU is caused by the cooling order on the blade.
However, this aspect has not been analyzed in more detail since the temperature does not significantly impact the performance, except for the outliers described in \secref{sec:perf_analysis}. 

\section{Impact of Performance Variations at Scale}
\label{sec:impact}

While the performance variations detailed above do -- to a large extent -- not affect codes that exhibit low scalability and thus do not scale across many nodes, these variations can have an impact on massively scalable codes that make use of several thousands of processor cores in parallel. 
For tightly coupled codes, a 10\,\% difference in performance between the worst and the best performing processor can mean that a significant amount of CPU cycles is spent on synchronization primitives.
While in the past, these losses were mainly caused by software-induced imbalances and were primarily a characteristic of the computational model and its implementation, the hardware can now induce significant imbalances itself. 
This constitutes an important shift for application developers: even perfectly balanced models may now run imbalanced, losing CPU cycles to hardware insufficiencies that are potentially unknown to and uncontrollable for the user.

The effects of performance variations on performance analysis can be tremendous. 
With previous processor generations, single runs were sufficient to determine imbalances in the code and to find hot spots that need optimization.
The new performance variations require additional information to distinguish between software- and hardware-induced imbalances and performance variations. 
This either requires incorporating additional information on the hardware by default (e.g., cycles consumed, actual frequency) or performing multiple runs to create a statistical performance model. 
Another solution would be the allocation of a fixed set of processors during the optimization process or to ensure the same configuration for every run, both of which can be either cost-ineffective or time-consuming due to the way HPC resources are allocated in production environments.

In order to estimate the effect of performance tuning actions, especially on a system-wide scale, the same holds true as with performance analysis: depending on the type of optimization employed, the effects might not immediately be visible but may be overshadowed by hardware-induced imbalances. 
The same limitations and mitigation strategies apply as for the performance analysis scenario described above. 
In both cases it is clear that execution time alone might not be a sufficient metric anymore since additional influence factors are added that are not under the control of the application developer or the performance analyst.

Several measures can be employed to mitigate the effects of the performance variations. 
During machine setup, the performance characteristics of all processors in the system can be determined and the information exposed to the job scheduler, employing a bonus or malus to ensure fair CPU time accounting. 
However, this does not work for applications scaling to the full system and might conflict with other allocation strategies, e.g., allocation of neighboring nodes for best interconnect performance. 

For applications that do not employ load balancing schemes but exhibit predictable imbalances, scheduling of heavy-loaded processes onto well-performing processors and vice versa can provide implicit rebalancing. 
However, this scheme requires a good understanding of the target application characteristics or an iterative process to provide information on predictable imbalances to the scheduler.

\section{Conclusion and Outlook}
This paper presents a survey of more than one thousand Haswell-EP processors, correlating their achievable performance and consumed average power during LINPACK runs and comparing the results with several hundred Sandy Bridge-EP CPUs. 
Power consumption readings are provided by both RAPL and the HDEEM measurement infrastructure. 
The resulting data indicates a shift in processor design from homogeneous performance characteristics to a TDP-bound, static power consumption with variable computational performance across the set of Haswell-EP processors under test.

The impact of this observation is significant on many levels of the HPC ecosystem.
One aspect is the serious impact on performance analysis, optimization, and tuning considering that any effects of code changes can be overshadowed by the processor characteristics, requiring careful measurements by the user.
It also makes performance modelling and prediction more complex.

Another effect on productive HPC usage is that formerly balanced, tightly coupled parallel applications may run imbalanced due to the heterogeneity of the cluster.
We discussed different approaches to mitigate this behavior.
Subclustering addresses this on a system-operations level to transparently provide heterogeneous subsets of compute nodes to applications.
Unbalanced applications can also make use of the heterogeneity by assigning tasks with more load to faster processors.
%
However, it remains as future work to investigate the impact of the performance variations on Haswell-EP and following generations on different workload types, e.g., memory-bound and real-world applications.

\section*{Acknowledgements}
This work has been funded in a part by the German Research Foundation (DFG) in the Collaborative Research Center ``Highly Adaptive Energy-Efficient Computing'' (HAEC, SFB 912) and by the European Union's Horizon~2020 Programme in the READEX project under grant agreement number 671657.


\bibliographystyle{splncs03}
\bibliography{paper}

\end{document}